\documentstyle [epsf]{mn}
\def\be{\begin{equation}}
\def\ee{\end{equation}}
\def\ba{\begin{eqnarray}}
\def\ea{\end{eqnarray}}

\def\ps{{P\&S }}

\begin{document}

\title[Assembly of Matter]
{\bf The Assembly of Matter in Galaxy Clusters}
\author[G.Tormen]
{Giuseppe Tormen \\
Max Planck Institute f\"{u}r Astrophysik,
Karl-Schwarzschild-Strasse 1, 85740 Garching bei M\"{u}nchen - GERMANY
and\\
Department of Physics, University of Durham, South Road, Durham
DH1 3LE England, UK\\
\smallskip
Email: bepi@mpa-garching.mpg.de}

\pubyear{1998}
\date{Accepted for publication in MNRAS}

\maketitle
\begin{abstract}
We study the merging history of dark matter haloes that end up in
rich clusters, using $N$-body simulations of a scale-free universe.
We compare the predictions of the extended Press \& Schechter (\ps)
formalism (Bond et al. 1991; Bower 1991; Lacey \& Cole 1993) with
several conditional statistics of the proto-cluster matter:
the mass distribution and relative abundance of progenitor haloes at 
different redshifts, the infall rate of progenitors within the 
proto-cluster, the formation redshift of the most massive cluster 
progenitor, and the accretion rates of other haloes onto it.
The high quality of our simulations allows an unprecedented 
resolution in the mass range of the studied distributions.
We also present the global mass function for the same 
cosmological model.
We find that the \ps formalism and its extensions cannot simultaneously 
describe the global evolution of clustering and its evolution in a 
proto-cluster environment. The best-fit \ps model for the global
mass function is a poor fit to the statistics of cluster progenitors. 
This discrepancy is in the sense of underpredicting the number of 
high-mass progenitors at high redshift.
Although the \ps formalism can provide a good qualitative description of 
the global evolution of hierarchical clustering, particular attention is 
needed when applying the theory to the mass distribution of progenitor
objects at high redshift.

\end{abstract}


\section{Introduction}\label{sec:intro}

The Press-Schechter formalism (Press \& Schechter 1974, hereafter \ps)
and its extensions (Bond et al., 1991; Bower 1991; Lacey \& Cole 1993,
hereafter LC93) provide a nice framework of analytical predictions for
the formation of dark matter haloes in cosmological models of hierarchical 
clustering. Until now, these predictions have given us the best handle we 
possess to understand the formation of structure in the real universe,
from sub-galactic scales up to the scale of galaxy clusters. 
It is therefore of great importance to know how reliable this model 
is, and how it fares when compared to numerical simulations of 
structure formation.

Although the \ps predictions have been extensively tested against 
$N$-body results (e.g. Efstathiou et al. 1988; Lacey \& Cole 1994,
hereafter LC94; Gelb \& Bertschinger 1994), numerical techniques are
only now reaching a resolution sufficient to extend this comparison
to a large range of masses. Due also to this limitation, most studies 
have been confined to the global mass function, while the 
predicted properties of halo progenitors have been compared to 
numerical results only by LC94.

A good understanding of the model and reliability for these 
{\em conditional} statistics is needed in order to reliably predict 
several observables, like 
the abundance of progenitors of any class of astrophysical objects, 
their formation times, their merging rates, and many other properties. 
These distributions are particularly important for semi-analytical 
models of galaxy formation (Kauffmann, White \& Guiderdoni 1993; Cole 
et al. 1994), where the extended \ps formalism is used to produce 
Monte Carlo realizations of the merging histories of dark matter 
haloes, and these in turn are the starting ingredient for producing 
galaxy populations in different cosmological models. 
The evolution of the properties of these populations changes
substantially for different choices of parameters in the \ps model.

In this paper we make a detailed analysis of the \ps predictions
for the formation of rich galaxy clusters, and compare these to 
the properties measured in $N$-body simulations of a scale-free
Einstein-de Sitter universe.
The high mass and force resolution of our simulations is ideal to 
resolve dark matter haloes over a large mass range.
Our result confirms that the \ps theory can be tuned to give a good 
overall fit of the global mass function. However, we will
prove that the \ps model best fitting the global mass function
of our simulations is not a good fit to the statistics involving 
the progenitors of clusters found {\em in the same simulations}.
This result points to a structural failure of the theory, as
the model parameter -- namely the collapse threshold $\delta_c$,
defined below in Section~\ref{sec:umf1} -- is fixed by the match
with the global mass function and cannot be changed when turning 
to the conditional statistics.

In Section~2 we briefly present the $N$-body simulations used in our 
study, and describe the method used to define their population of 
dark matter haloes.
In Section~3 we calibrate the \ps model by fitting the global 
mass function of a set of cosmological simulations. We 
show that the standard value for the collapse threshold
$\delta_c=1.69$ provides a good fit to the simulations.
In Sections~4 and 5 we consider a sample of clusters extracted from the
simulations presented in Section~3: we present several conditional 
statistics of the cluster progenitors, and compare them with the 
predictions of the extended \ps formalism.
We show that the \ps model, calibrated as in Section~3, is not a good 
description of the evolution of cluster progenitors.
Finally, in Section 6 we discuss our results and summarize the main 
conclusions of the paper.

\section{Method}\label{sec:method}

\subsection{The Simulations}\label{sec:sim}

The cluster $N$-body simulations studied in this paper have been
presented in Tormen, Bouchet \& White (1997), where complete 
details may be found. 
In summary, our sample consists of nine dark matter haloes of rich 
galaxy clusters. These have been produced by tailor-made, 
high-resolution $N$-body simulations, using the initial condition 
resampling technique described in Tormen et al. (1997).
The clusters correspond to the nine most massive objects found in a 
{\em parent} cosmological simulation (White 1994) of an Einstein--de 
Sitter universe, with scale-free power spectrum of fluctuations 
$P(k) \propto k^n$, and a spectral index $n=-1$, the appropriate 
value to mimic a standard cold dark matter spectrum on scales 
relevant to cluster formation. 
All simulations have a Hubble parameter $H_0 = 50$ km s$^{-1}$ Mpc$^{-1}$,
and are normalized to match the observed local abundance of galaxy 
clusters (White et al. 1993).

The parent simulation was evolved using a Particle-Particle-Particle-Mesh 
code (Efstathiou et al. 1985) with $100^3$ particles in a $256^3$ 
grid with periodic boundary conditions. The box size is 150 Mpc 
on a side. 
The cluster simulations were evolved using a binary tree-code,
in proper coordinates, on a sphere of diameter 150 Mpc, with vacuum
boundary conditions.
The average cluster mass over our high-resolution sample is 
$M_v \simeq 1.1 \times 10^{15} M_{\sun}$. Each cluster is resolved 
by $\approx 20000$ dark matter particles with an effective force 
resolution of $\sim 25$ kpc which is constant in proper coordinates.

\subsection{Identification of dark matter lumps}\label{sec:hid}

We used the potential energy of particles to identify the centres
of dark matter lumps (Efstathiou et al. 1988).
We defined lumps by an overdensity criterion, and included all particles 
within a sphere of mean overdensity $\delta_v = 178$, centred on the 
particle with lowest potential energy. The value $\delta_v$ corresponds
to the virial overdensity in the model of a spherical top-hat collapse
in an Einstein-de Sitter universe.
We will call the corresponding radius the virial radius $r_v$ of the lump.

\begin{figure*}
\epsfxsize=\hsize\epsffile{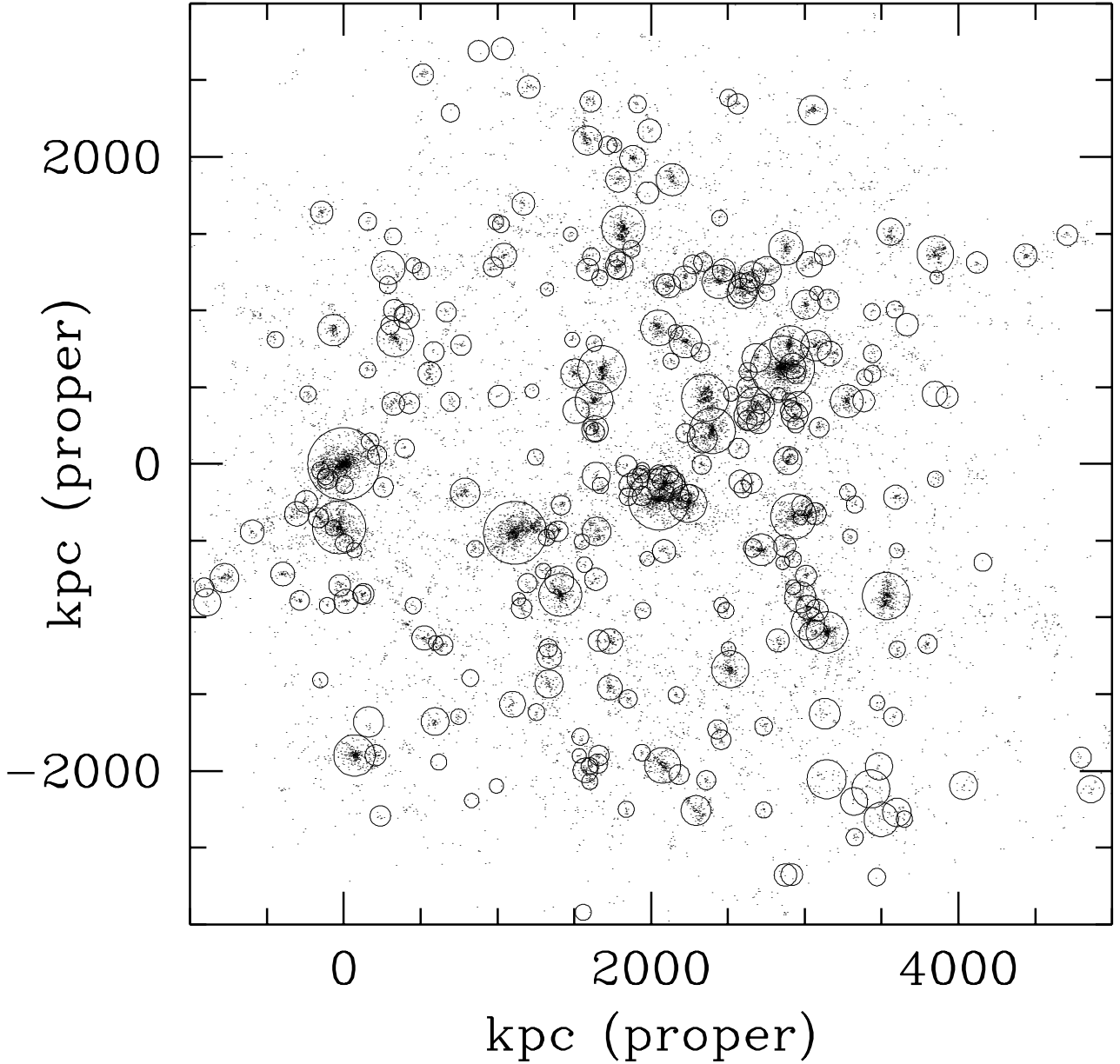}
\caption{The figure shows the progenitor haloes of a cluster in 
our sample, seen at redshift $z = 2.8$; the cluster virial mass 
at redshift $z = 0$ is $M_v = 7 \times 10^{14} M_{\sun}$.
The circles show the lumps of matter found by our algorithm, based 
on local minima of gravitational potential. The radius of each
circle is the virial radius of the lump. Lumps contain at least 8
particles. Overlapping circles are only due to projection effects. 
The most massive cluster progenitor is situated at position (0,0), 
and is formed by $\approx 1300$ particles. Only particles actually
ending up in the final cluster are plotted. Therefore, progenitor 
haloes which carry only a small fraction of their mass into the 
final cluster have only few particles drawn within the corresponding 
circles.}
\label{fig:circles}
\end{figure*}

Following Tormen (1997) we limit our analysis to haloes with 
$n_v \geq 8$ particles. We will generally call {\em field particles} 
all particles unclustered or in lumps with $n_v < 8$.
Fig.~\ref{fig:circles} shows an example of the lumps found by this 
algorithm. The plot refers to a cluster in our sample, observed at
redshift $z = 2.8$. Only particles which are progenitors of the final
cluster are plotted.

\section{Calibration of the \ps model}
\label{sec:cal}

In this Section we calibrate the \ps model by fitting the predicted
global mass function of dark matter haloes to the results of our 
cosmological simulations. We take the collapse threshold
$\delta_c$ as a free parameter and determine the value best-fitting
the numerical data. 
We assume a top-hat filtering to define the mass variance $\sigma^2(R)$.
In this Section, and only here, we use two methods to define the 
haloes. A standard percolation scheme -- the friend-of-friends 
algorithm (hereafter FOF) -- as defined in Davis et al. (1985), and 
a spherical overdensity criterion (herafter SO), as defined in the 
previous Section. We do this to demonstrate the robustness of our
best fitting results. Having proven that, in the remainder of the
paper we will only consider SO haloes.

For the FOF we take a linking length $b=0.2$. With this choice, 
groups are approximately identified by closed surfaces of local 
density $\rho/\rho_b \approx 1/b^3 = 125$ (factors of 2 or 1/2 
may apply, see Frenk et al. 1988; LC94). 
Following the notation of LC94, we will call 'FOF(0.2) haloes' the 
groups of particles found by this algorithm. For the SO haloes we 
take the virial overdensity $\delta_v = 178$, and will call them 
'SO(178) haloes'. As discussed in LC94, the algorithms we use to define 
the dark matter haloes are essentially equivalent to those used by LC94.

\subsection{Global Mass function}\label{sec:umf1}

The study of the global mass function requires a full 
cosmological simulation. Taking advantage of the self-similarity 
of our model, we present results averaged over ten output 
times, taken from two different simulations: one is the parent 
of the cluster simulations presented above, the other is identical 
to the first in the numerical parameters, and only differs in the 
realization of the initial conditions.
An extensive comparison between the predictions of the \ps formalism 
and the results of scale-free $N$-body simulations has been performed
by LC94. At the end of this Section we will compare our results to 
theirs.

For an Einstein-de Sitter universe with scale-free power spectrum
$P(k) \propto k^n$, the fraction of mass in haloes with mass $M$
at redshift $z$, per interval $d\ln M$, is predicted by the \ps
formalism to be
\ba
{df_M \over d\ln M}(M,z) &=& {\alpha \over 2}
\left({2 \over \pi} \right)^{1/2}
\left(  M \over M_*\right)^{\alpha/2} \nonumber \\
&\times& \exp\left[-{1 \over 2} \left({M \over M_*} \right)^\alpha\right],
\label{eq:udmf}
\ea
where $\alpha = (n+3)/3$. The time dependence enters only through 
the characteristic non-linear mass $M_*(z) = (4\pi/3)R_*^3 \rho_b(z)$,
with $\rho_b(z)$ the background density of the universe at redshift
$z$, and $R_*(z)$ the scale corresponding to a linear overdensity 
$\delta_c$ of order unity. The specific value of $\delta_c$, called 
{\em collapse threshold}, depends on the criterion used to identify 
dark matter haloes; a sensible starting value is $\delta_c = 1.69$, 
the linear extrapolation at collapse for a spherical top-hat. 
We use a top-hat filter to calculate $M_*$. The value of $M_*$ for 
$\delta_c = 1.69$ is $M_*|_{z=0} = 6.16\times 10^{13} M_{\sun}$.

The goal of this Section is to determine the value of $\delta_c$
(that is, of $M_*$) for which Eq.~(\ref{eq:udmf}) best fits the 
mass function of haloes from simulations.

\begin{figure*}
\epsfxsize=\hsize\epsffile{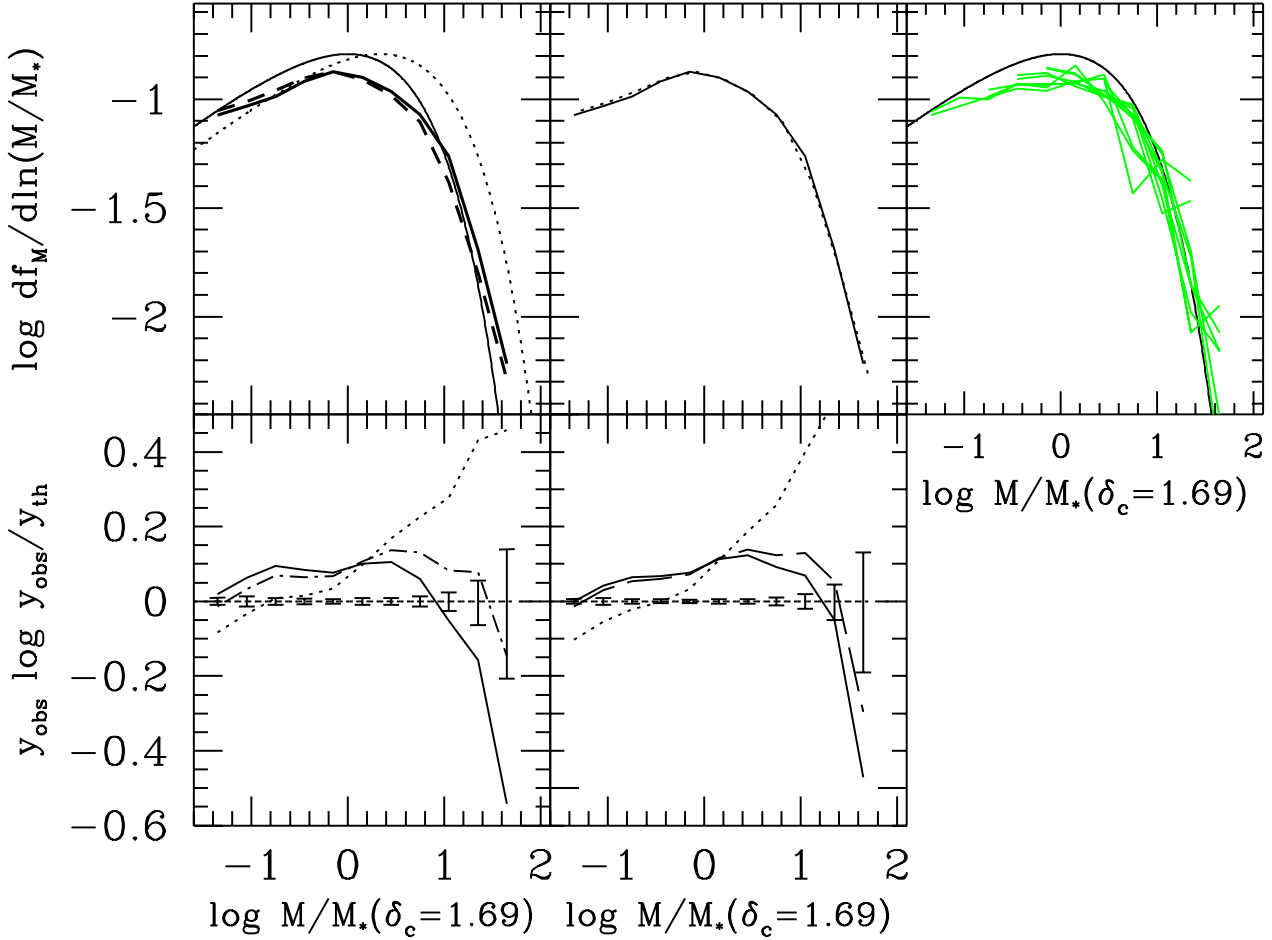}
\caption{Global mass function for the scale-free cosmological
model with $n=-1$.
{\bf Top Left:} Thick curves indicate the values found in simulations,
for different group finding algorithm: FOF(0.2) (solid) and SO(178) 
(short dash). Smooth curves show the values predicted by 
Eq.~(\ref{eq:udmf}) for different values of $\delta_c$
{\bf Top center:} The SO(178) curve is shifted by 0.065 along the
horizontal axis, showing that its shape is almost identical to the
FOF(0.2) curve.
{\bf Top right:} Mass function for 10 different output times (broken
curves), taken from two different $N$-body runs of the same model.
All outputs agree with the \ps prediction with $\delta_c=1.69$
(smooth curve), and there is no systematic variations between the
two runs or between different time outputs. This result is shown for 
FOF(0.2) groups, but identical results are obtained for SO(178) groups.
{\bf Bottom Left:} Logarithmic residuals from the top left panel.
for FOF(0.2) groups. Error bars represent the Poissonian $1\sigma$ 
of the mean around the observed value.
{\bf Bottom Right:} Corresponding residuals for SO(178) groups.
Different line types in all panels correspond to different values 
of $\delta_c$: 1.3 (dotted), 1.55 (dot-dash), 1.63 (long dash) and 
1.69 (solid).}
\label{fig:udmf}
\end{figure*}

Fig.~\ref{fig:udmf} presents the global mass function 
$df_M/d\ln M$ for the scale-free model with $n=-1$. The average 
over different output times was done as described in LC94, weighting 
the contribution of each curve in each bin by the number of haloes 
in that bin. To facilitate comparison with LC94, we only used haloes 
with at least 20 particles.

The top left panel shows the logarithm of $df_M/d\ln M$ for FOF(0.2) 
groups (thick solid), and for SO(178) groups (thick dashed), together 
with the prediction of Eq.~(\ref{eq:udmf}) for different values of 
collapse threshold: the canonical value $\delta_c = 1.69$ (thin solid) 
and $\delta_c=1.3$ (thin dotted). The choice of the latter value 
is explained below.

From the figure we see that the mass function from simulations is 
globally flatter than the \ps prediction, especially at the high mass 
end. This finding is however in general agreement with previous studies
(e.g. Efstathiou et al. 1988, LC94). Despite the slight difference
of shape, the $N$-body results are reasonably fitted by the \ps prediction
with the canonical choice $\delta_c = 1.69$, both for FOF(0.2) and
SO(178) groups. In fact, the agreement between model and simulation
curve in the mass range $-1.4 \la \log_{10}(M/M_*) \la 1.4$ is better
than $\approx 40$ percent for FOF(0.2) haloes and better than 
$\approx 35$ percent for SO(178) haloes.
As the shape of the curves from $N$-body is slightly different from 
the predicted one, best fitting is a rather subjective exercise. 
However, with the choice $\delta_c=1.55$ for FOF(0.2) haloes one can 
slightly improve the fit at high masses, up to 
$\log_{10}(M/M_*) \simeq 1.7$.

The figure also shows that FOF(0.2) groups are slightly more massive
than SO(178) groups. In fact, if we shift the SO(178) curve horizontally
by 0.065 in $\log_{10}(M/M_*)$, as in the top central panel, the two curves 
match almost exactly. Using this scaling between masses of haloes
defined by the two algorithms, the best fitting value $\delta_c=1.55$ 
for FOF(0.2) haloes correspond to $\delta_c=1.63$ for SO(178) haloes 
respectively. The residuals between model and $N$-body data, for 
the considered values of $\delta_c$, are plotted in the lower panels 
of Fig.~\ref{fig:udmf}.

The top right panel shows the 10 individual curves used to define the
average mass function shown in the top left panel.
Note that all outputs agree with the \ps prediction with $\delta_c=1.69$
and there is no systematic variations between the two runs or between 
different time outputs.

\subsection{Discussion}\label{sec:umf2}

Let us compare in more detail our results with those of LC94. These
are based on scale-free simulations, one for each spectral index of 
fluctuations $n=-2$, $n=-1$, $n=0$. LC94 found that the value of 
$\delta_c$ best fitting the mass function of their simulations is 
$\delta_c = 1.96$ for the SO(178) haloes and $\delta_c = 1.81$ for 
the FOF(0.2) ones. With this choice, they could fit the $N$-body 
results to better than 30 per cent over a mass range that, for the 
$n=-1$ simulation, corresponds to $-1.5 \la \log_{10}M/M_* \la 1.4$.
They also found that the canonical value $\delta_c = 1.69$ still
provides a reasonable fit, with an accuracy better than a factor 
of two on the same mass range for the FOF(0.2) groups.

Our results are generally in good agreement with LC94, as we 
also find that {\em (a)} SO(178) haloes are slightly smaller 
than FOF(0.2) haloes, and {\em (b)} $\delta_c = 1.69$ is an
overall good fit to the data for groups in both identification
schemes. Some minor differences are due to the fact that our 
results are only based on the $n=-1$ model, as opposed to the 
three scale-free models used by LC94. 

Therefore, our robust conclusion is that the \ps model with 
$\delta_c = 1.69$ is an appropriate description of the clustering 
of matter in a global cosmological environment, for both SO(178) 
and FOF(0.2) dark matter haloes. In the remainder of the paper we 
will compare the predictions of this {\em calibrated} model to the 
conditional statistics of the progenitors of the most massive 
clusters found in one of the two simulations presented above.

Another result is that the choice $\delta_c = 1.3$ gives
clearly a bad fit to the global mass function (long
dashed lines in Fig.~\ref{fig:udmf}), as it predicts far
too many high mass haloes. This result is important, as we
will show below that, incidently, $\delta_c = 1.3$ would be 
the value appropriate to describe the conditional statistics 
related to the cluster progenitors.

Finally, let us comment on the relative mass of FOF(0.2) and 
SO(178) haloes. As we mentioned before, haloes identified with 
a FOF criterion are approximately bounded by a surface of local 
overdensity $\rho/\rho_b \approx 125$ for $b=0.2$.
On the other hand, SO(178) haloes are cut at a mean overdensity
$<\rho>/\rho_b = 178$. 
Since we know the shape of the density profiles for these haloes 
(Navarro, Frenk \& White 1996; Cole \& Lacey 1996; Tormen et al.
1997), we can relate these two values.
The virial mean overdensity corresponds to a local overdensity 
$\rho/\rho_b \approx 10 - 30$, weakly depending on the halo mass 
in terms of $M_*$ and on the cosmological model. 
On the other hand, a local overdensity $\rho/\rho_b \approx 125$ 
corresponds to a mean overdensity $<\rho>/\rho_b = 500$ (as found 
also by Frenk et al. 1988), and to a mass 
$M_{FOF(0.2)} \approx 0.75 M_{SO(178)}$, with $M_{SO(178)}$ the 
virial mass. From this information, we should expect the FOF(0.2) 
groups to be less massive, not more massive than the SO(178) haloes. 
Instead we have shown above that, on average, 
$M_{FOF(0.2)} \approx 1.15 M_{SO(178)}$, or about 50 per cent more
massive than their expected mass. This difference is probably due 
to the tendency of FOF to connect structures along bridges and 
filaments, and is a quantitative evidence that FOF groups are, 
on average, very aspherical objects.

\section{Evolution of cluster progenitors}\label{sec:mass}

We now turn our analysis to the cluster simulations. We remind that
the clusters analyzed here have been obtained by resimulating at
higher mass and spatial resolutions the nine most massive clusters
found in one of the two $N$-body runs presented in
Section~\ref{sec:umf1}.
In this Section we study the mass function of the cluster progenitors
at different redshifts, the evolution of the comoving density 
of progenitors, and their global infall rate as haloes merge 
together to form more massive haloes.
In the next Section we will concentrate on the most massive
cluster progenitor: we will study its formation redshift and
the time evolution of the corresponding infall rate.
We will prove that the calibrated \ps model (which has
$\delta_c=1.69$), underpredicts the clustering of high-mass
progenitors at high redshift. As mentioned above, in the remainder 
of this paper we only consider haloes defined by the spherical 
overdensity criterion SO(178).

In the hierarchical clustering picture of structure formation, 
matter clusters on small scales first, and haloes of a given mass 
are formed by the assembly of pre-existing smaller haloes.
Therefore, in this model, once an element 
of matter becomes part of a halo, it will be forever part of some 
halo from then on. Also, all the matter forming a halo comes from 
haloes of some mass. Finally, when a satellite lump is accreted 
onto a more massive halo, all of the satellite mass is incorporated 
in the more massive object. In this picture the formation of structure 
is a very lumpy process. 
This description is fairly close to what we observe in $N$-body 
simulations of hierarchical clustering, although the actual collapse 
of structures is more complicated than its idealized version; 
in particular, satellites may be tidally stripped prior to merger
with a more massive haloes, so that only a fraction of the satellite 
mass becomes part of the accreting system. 
Also, during violent merging events, energy is re-distributed 
among the mass elements, and some matter may temporarily exit 
the virial radius of the newly formed halo, or in principle even 
become unbound. 

\begin{figure}
\epsfxsize=\hsize\epsffile{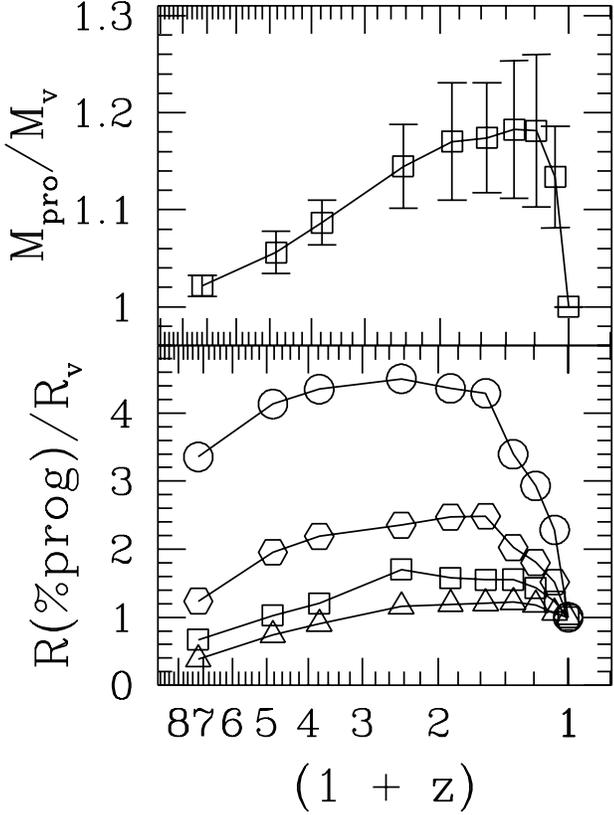}
\caption{{\bf Top panel:} fraction of the total mass in progenitors
which will actually end up within the virial radius of the final 
cluster. {\bf Bottom panel:} radius of the sphere centered on the 
final cluster and enclosing $90\%$, $95\%$, $99\%$, and $100\%$ of 
the total mass in progenitors.}
\label{fig:tmpro}
\end{figure}

Let us call, at any given redshift, {\em progenitors} of a cluster 
all haloes containing at least one particle that by $z=0$ will be
part of the cluster, and also all field particles which will end 
up in the final cluster. We can ask: does all the mass of the 
progenitors end up in the final system? 
To answer we calculated the ratio $M_{pro}(z)/M_v$ of the total mass 
in progenitors (both in haloes and in field particles) at redshift 
$z$, to the final cluster mass. This is shown in the upper panel 
of Fig.~\ref{fig:tmpro}. Initially, before the formation of 
any halo, this value is unity. Then, as matter starts to collapse 
into haloes, progenitor and non-progenitor particles are captured 
in the same haloes, and $M_{pro}(z)/M_v$ increases up to $\simeq 1.2$ 
at $z\sim 1$. It then remains at his level until $z \simeq 0.2$, then
it decreases to reach unity at the final time.

What is the fate of the non-progenitor particles captured in 
progenitor haloes? That is, where can we find, at $z = 0$, the 
extra $\approx 20\%$ of matter in progenitors shown in the upper 
panel of Fig.~\ref{fig:tmpro}? The answer is shown in the lower
panel, where we plot the radius of spheres centered on the final
cluster and containing different percentages of the total progenitor
mass, as identified at different redshifts. For example, we see that
99 per cent of the progenitor particles is situated, at $z = 0$, 
within roughly 2.5 virial radii, and that 100 per cent of them is 
within 4.5 virial radii. This means that, during the cluster collapse, 
no particle from the progenitor haloes has actually escaped to infinity.
Correspondingly, but implicitly required by the re-simulation technique
used for the cluster simulations, no particle from infinity has become
part of the final halo.
By number, roughly $80\%$ of the progenitor lumps carry at least
$80\%$ of their mass into the final cluster, while only $10\%$
carry less than $20\%$ of their mass, in fair agreement with the 
theoretical picture.

A final remark is on resolution issues. The gravitational softening 
of our simulations is constant in proper coordinates. Since at high 
redshift haloes of a given mass are smaller, they are also more poorly
resolved. As a softening comparable to the halo size introduces a bias
in the halo mass, in this section we conservatively consider only 
progenitor haloes with virial radius larger than $100$ kpc, or four 
times the gravitational softening. This will result in a time
dependent low mass cutoff in the statistics we are going to present.

\subsection{Conditional mass function}\label{sec:cmf}

\begin{figure*}
\epsfxsize=\hsize\epsffile{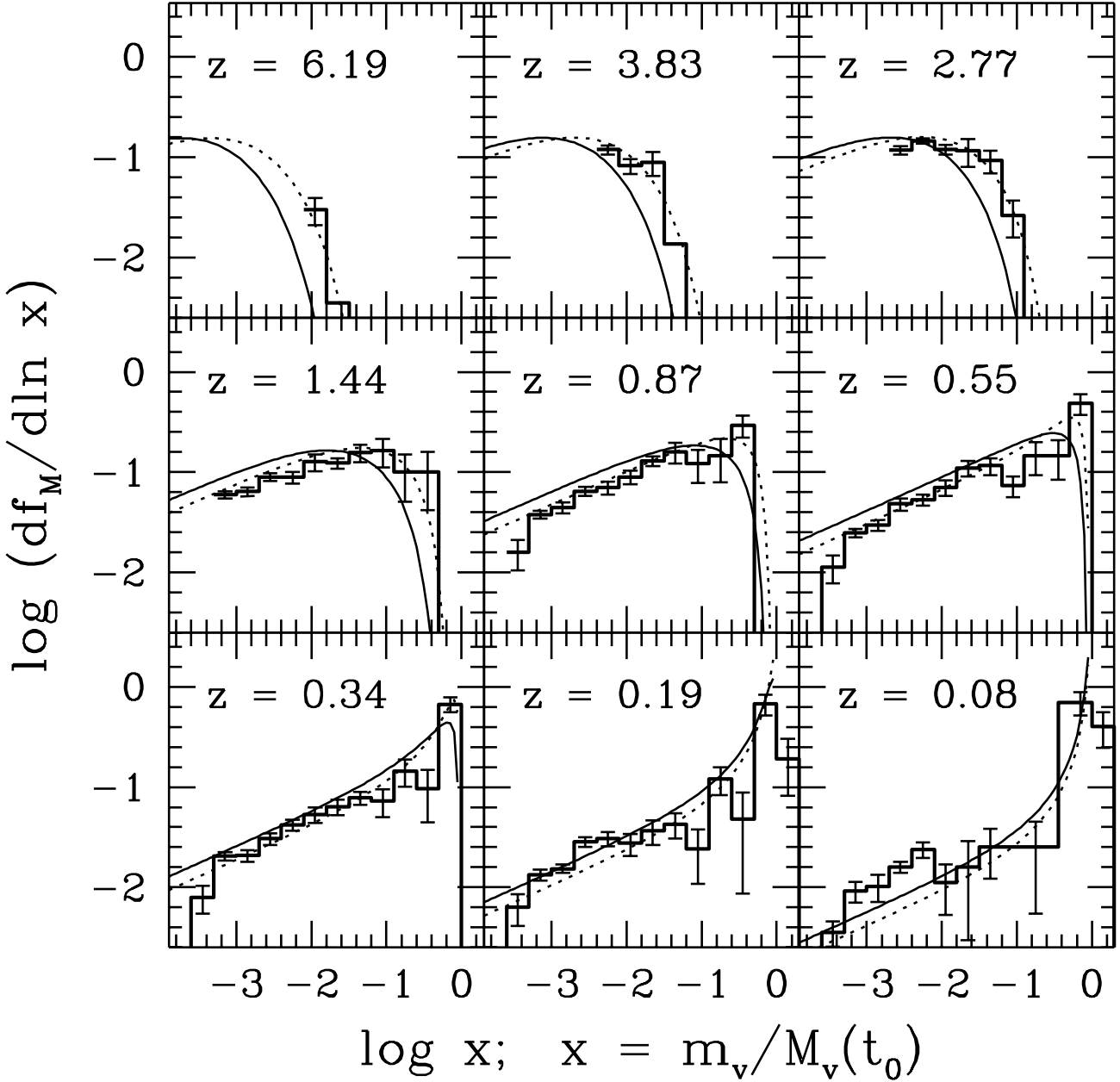}
\caption{Differential mass function progenitors at different
redshifts. Each panel shows the mass function of the cluster 
progenitors, averaged over the cluster sample. Error bars are 
$1\sigma$ of the distribution, calculated from 100 bootstrap 
resamples of the actual sample. 
The smooth solid curve represents the calibrated \ps prediction
(Eq.(\ref{eq:cdmf}) with $\delta_c = 1.69$).
The smooth dotted curve is a \ps model with $\delta_c = 1.3$.}
\label{fig:cdmf}
\end{figure*}

\begin{figure*}
\epsfxsize=\hsize\epsffile{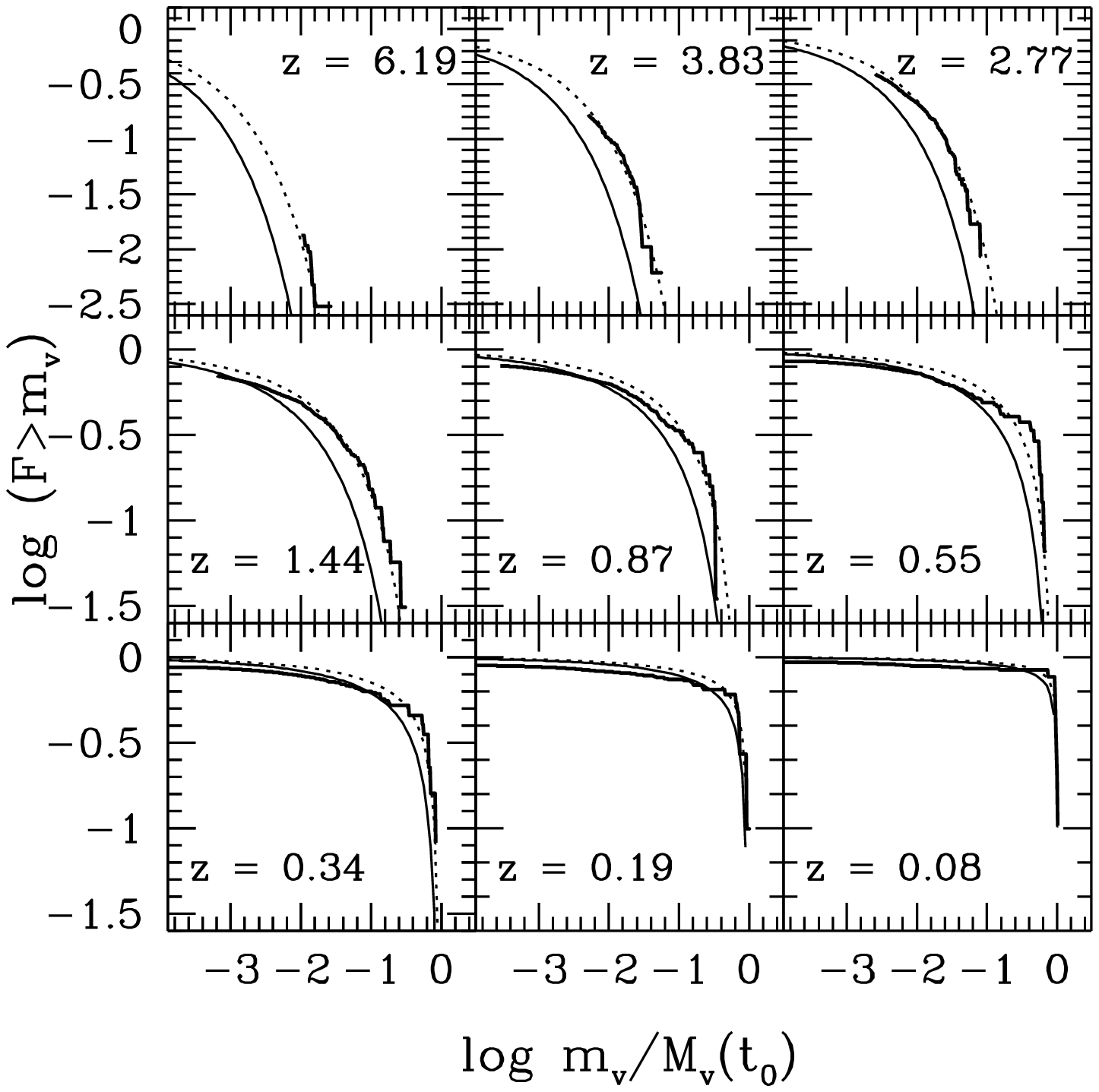}
\caption{Cumulative mass function of progenitors at different
redshifts. The thick solid curves are the simulation result.
The thin smooth curves show the predictions of Eq.(\ref{eq:ccmf}), 
with $\delta_c = 1.69$ (solid line) and $\delta_c = 1.3$ (dotted 
line) respectively.}
\label{fig:ccmf}
\end{figure*}

In this section we fit the calibrated \ps model
to the conditional mass function of cluster progenitors.
In the \ps formalism, the mass fraction in progenitor haloes of a 
given mass can be easily calculated. Eq.(2.15) of LC93 gives the 
differential mass distribution $df_M(M,z | M_2, z_2)/d\ln M$ of 
haloes at redshift $z$, conditioned by the requirement that the 
matter will be in a cluster of given mass $M_2$ at redshift 
$z_2 < z$. For cosmological models with scale-free power spectrum 
of fluctuations: $P(k) \propto k^n$, the equation reads
\ba
{df_M \over d\ln M}(M,z | M_2,z_2) &=&
\alpha (2\pi)^{-1/2}
\left(
  M \over M_*
\right)^{\alpha/2}
{(z - z_2) \over (1 - x^{\alpha})^{3/2}} \nonumber \\
&\times&
\exp
\left[
  {-(z - z_2)^2 \over 2 (1 - x^{\alpha})}
  \left(
    {M \over M_*}
  \right)^\alpha
\right]
\label{eq:cdmf}
\ea
where $x = M/M_2$. In our case we take $z_2 = 0$ as the present time 
in Eq.(\ref{eq:cdmf}), and $M_2 = M_v$ as the final cluster mass.
The fraction of mass in haloes with mass larger than $m_v$ at redshift 
$z$ is easily obtained by integrating Eq.(\ref{eq:cdmf}) over M, from
$m_v$ to infinity. The resulting cumulative mass function is
\ba
f_M(M>m_v, z | M_v) &=& {\mathrm erfc} 
\Bigl\{
  {z \over \sqrt{2}}
  \left(
    {m_v \over M_*} 
  \right)^{\alpha / 2} \nonumber \\
  &\times&
  \left[
    1 - 
    \left(
      {m_v \over M_v}
    \right)^\alpha
  \right]
  ^{-1/ 2}
\Bigr\}.
\label{eq:ccmf}
\ea
In Figs~\ref{fig:cdmf} and \ref{fig:ccmf} we compare the 
differential and cumulative mass function of progenitors 
Eq.(\ref{eq:cdmf}) and Eq.(\ref{eq:ccmf}) to the corresponding
statistics from our simulations.
Abscissa are normalized to the final mass of the cluster. 

Data from all nine clusters are weighted by the number of particles 
in progenitors, and combined together. Each panel refers to a 
different time output, labeled by the average redshift over the sample.
For the differential mass function, error bars were obtained from 
$100$ bootstrap resampling of the data. That is, for each bin they 
indicate the $1\sigma$ dispersion of the distribution of 100
re-samples of 9 clusters each, each re-sample randomly drawn 
with replacement from the original collection of 9 clusters.
The total number of progenitors in each histogram varies from
over 2000 at high redshift to a few hundred close to $z=0$.
Histograms are built using the virial mass of the progenitor haloes.

It is clear from the two figures that the calibrated \ps model
(Eqq.~(\ref{eq:cdmf}), (\ref{eq:ccmf}), smooth solid curves)
is not a good fit to the numerical data. Specifically, it
underestimate the number of massive progenitors by up to one order of 
magnitude at redshift $z \ga 1$. 
Now, {\em it happens} that the simulation results are reasonably 
fit by a \ps model with $\delta_c = 1.3$ (corresponding to 
$M_*|_{z=0} = 1.35 \times 10^{14} M_{\sun}$, dotted curve).
However, the collapse threshold has been fixed by the fit to 
the global mass function and is not a free parameter anymore.
Therefore, the dotted curve must be considered only as a rough 
quantitative measure of the discrepancy between the calibrated 
\ps model (which has $\delta_c = 1.69$) and the simulations. 

In the present analysis we have used the total mass of the progenitor 
haloes. However we found that the result does not change if instead 
one uses, for each progenitor, the actual fraction of halo mass ending
in the final cluster.

\subsection{Comoving density of progenitors}\label{sec:hab}

Another way to look at the same results is to consider the comoving 
density $\rho_{pro}(z)$ of progenitor haloes in a given mass range
$[M_a,M_b]$, at different redshifts.
This is easily obtained as an integral over Eq.~(\ref{eq:cdmf}):
\ba
\rho_{pro}(M_a,M_b,z|M_2,z_2) &=& 
\rho_b\int_{\ln M_a}^{\ln M_b}  \nonumber \\
&& {df_M \over d\ln M}(M,z | M_2,z_2) {d\ln M \over M}.
\label{eq:hab}
\ea
The corresponding number of progenitors per cluster in the same mass 
range is $N_{pro}(z) = \rho_{pro}(z) M_2/\rho_b$.
This quantity is similar to the comoving density of haloes usually 
calculated from the global mass function Eq.~(\ref{eq:udmf}),
which is often compared to the observed abundance of some class of
extragalactic objects; it differs however in the interpretation. 
The comoving density Eq.~(\ref{eq:hab}) in fact refers to proto-cluster
matter, and so it only applies to regions of the universe initially
overdense, and unusually populated. This {\em conditional} density
can therefore be used to estimate the predicted comoving density
of objects which populate the environment of a proto-cluster.

\begin{figure}
\epsfxsize=\hsize\epsffile{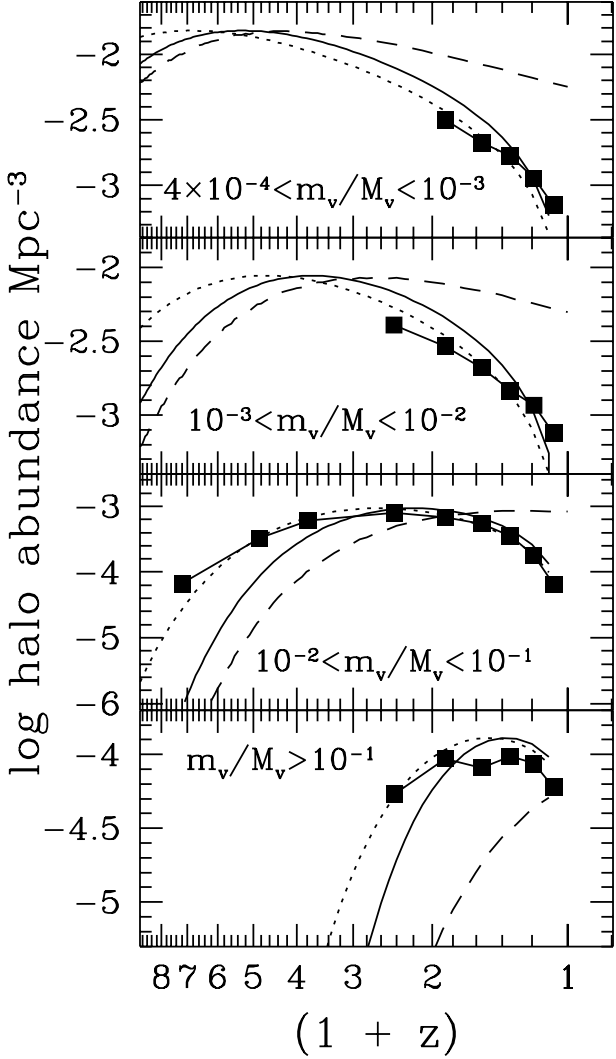}
\caption{Comoving density of progenitor haloes in different mass 
ranges, compared with the extended \ps prediction, versus redshift. 
The solid squares show the average number of haloes per comoving cubic 
Mpc, measured in the simulations. The smooth solid curve is the 
conditional calibrated \ps prediction Eq.~(\ref{eq:hab}). The 
dashed line is the corresponding prediction for an unconstrained 
environment, obtained by inserting Eq.~(\ref{eq:udmf}) in 
Eq.~(\ref{eq:hab}). The dotted curve is a conditional \ps prediction 
with $\delta_c = 1.3$.}
\label{fig:hab}
\end{figure}

In Fig.~\ref{fig:hab} we show comoving density for four different
mass ranges and for $z_2 = 0$. In each panel, the number of progenitors
per comoving cubic Mpc is plotted versus redshift: the symbols show 
the average value over the cluster sample. Virial masses for the 
progenitor SO(178) haloes are considered.

The solid curve is the conditional density predicted by the calibrated
\ps model. For comparison, the dashed line shows the corresponding 
global comoving density obtained by inserting the global mass
distribution Eq.(\ref{eq:udmf}) into Eq.~(\ref{eq:hab}).
The \ps model is for a final halo of mass 
$M_v = 1.1\times 10^{15} M_{\sun}$, the average cluster mass for our 
sample.
Again we see that the calibrated \ps model (solid curve) underestimates
the clustering of high-mass progenitor by up to two orders of 
magnitude at high redshift. Again, the discrepancy is roughly
described by a change in collapse threshold ($\delta_c=1.3$,
dotted curve).

\subsection{Global infall rates}\label{sec:irate}

A measure of the rate at which the proto-cluster matter merges
into more and more massive haloes is given by the time derivative, 
at any fixed mass, of the cumulative mass function Eq.(\ref{eq:ccmf}). 
The expression is (Bower 1991)
\ba
{\partial f(m_v,z|M_v) \over \partial t} 
& = & 
    {\partial \over \partial x} {\mathrm erfc}(x_M) 
    {\partial x \over \partial z}
    {\partial z \over \partial t} \nonumber \protect\\
& = & 
    \left(
      {2\over\pi} 
    \right)^{1/2} 
    {2 \over 3}(1 + z)^{5/2} 
    \left(
      {m_v \over M_*}
    \right)^{\alpha/2} \nonumber \protect\\
& \times &
    \left[1 - 
      \left(
        {m_v \over M_v} 
      \right)^\alpha 
    \right]^{-1/2} \nonumber \protect\\
& \times &
        \exp
        \Bigl\{ 
          {z^2 \over 2} 
          \left(
            {m_v \over M_*}
          \right)^\alpha
          \left[1 - 
            \left(
              {m_v \over M_v}
            \right)^\alpha
          \right]^{-1}
         \Bigr\},
\label{eq:irate}
\ea
with $x_M$ the content of the curled brackets in Eq.~({\ref{eq:ccmf}}).
More specifically, Eq.~(\ref{eq:irate}) gives the global rate at 
which progenitors with mass smaller than $m_v$ merge together and 
form progenitors of mass larger than $m_v$, at redshift $z$. 
This quantity has been called {\em infall rate} by Bower (1991).
Therefore, we keep this definition, although we will distinguish 
below between the global infall rate of progenitors, given by 
Eq.~(\ref{eq:irate}), and the infall rate of haloes onto the main 
cluster progenitor, which we calculate in Section \ref{sec:arate}.
It is in fact important to remember that Eq.~(\ref{eq:irate})
does not contain any information on the relative mass of the 
merging haloes, nor does it distinguish mergers onto the most 
massive cluster progenitor from other encounters between progenitors. 

\begin{figure*}
\epsfxsize=12.truecm\epsffile{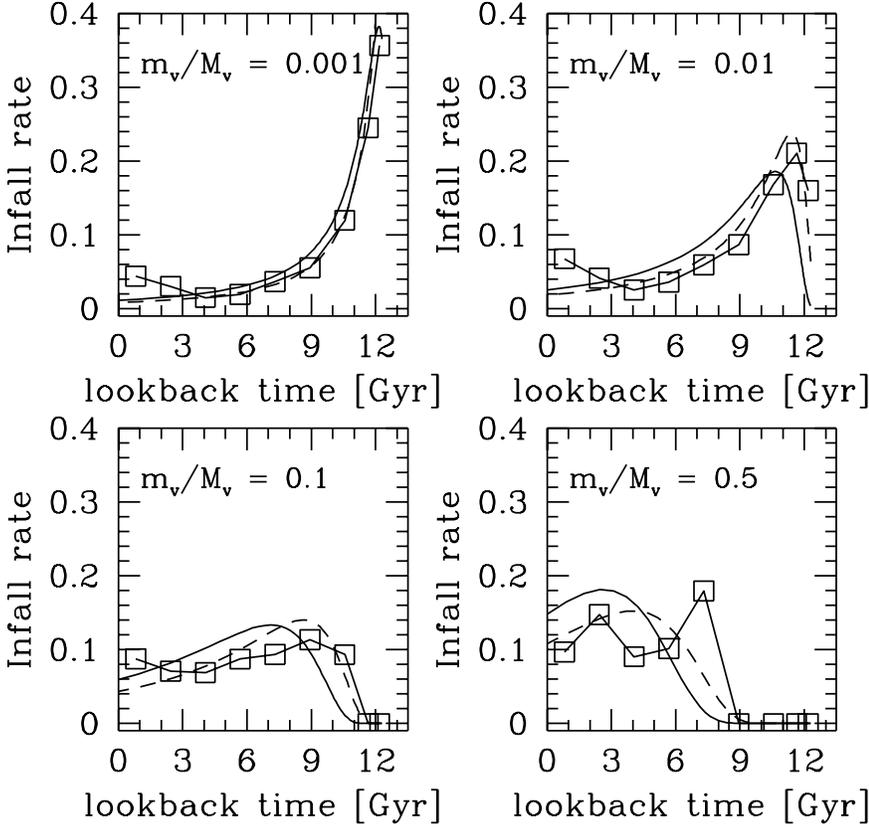}
\caption{The figure shows the global infall rates of cluster
progenitors, as a function of redshift, for different threshold
mass $m_v/M_v$.}
\label{fig:irate}
\end{figure*}

It is useful to plot the information of Eq.~(\ref{eq:irate}) as
a function of the lookback time from $z=0$.
This is done in Fig.~\ref{fig:irate}, where we compare the 
predictions of Eq.~(\ref{eq:irate}) with the infall rate measured 
from the simulations, taking the numerical derivative of
the curves shown in Fig.~\ref{fig:ccmf}. Rate units are fraction
of the final cluster mass per Gyr. 

Each panel shows the infall rate for a given mass threshold, in 
terms of the final cluster mass. The calibrated \ps prediction
(solid curve) is systematically shifted to lower lookback times
for $m_v/M_v>0.01$, compared to the numerical results.
The dotted curve is a \ps model with $\delta_c = 1.3$, shown
to quantify this discrepancy.

\section{Formation of the main progenitor}
\label{sec:mainp}

In the previous Section we have considered the global evolution of
all progenitors of a present-day cluster. A different way to estimate 
the speed at which the cluster grew is to look at the formation of the
most massive cluster progenitor, which we do in the present Section.
We will focus on the redshift at which the main progenitor formed, 
and on the infall rate of other cluster progenitors onto it.

Operationally, we define the {\em main} (or {\em largest}) 
{\em progenitor} of a cluster, at any given redshift, as the 
progenitor halo containing the largest fraction of the mass 
from the final cluster.
In all cases this coincided with the most massive progenitor,
although, with the present definition, the main cluster progenitor 
in $N$-body simulations is not necessarily the most massive one,
as progenitors carry different fraction of their mass into the
final cluster.

\subsection{Formation redshift}
\label{sec:ftime}

The cluster {\em formation redshift} is usually defined as
the earliest redshift at which the largest cluster progenitor 
reached at least $50\%$ of the cluster's final mass. 
The argument given by LC93 to estimate formation redshifts is the 
following. The probability for a cluster of present mass $M_2$ to 
have had at least a progenitor of mass $M$ at redshift $z$ is simply 
the probability in Eq.~(\ref{eq:cdmf}) times the average number $M_2/M$ 
of progenitors with mass $M$. 
Since each object can have at most one progenitor with half or more 
of its proper mass, the probability that a cluster had a progenitor 
with mass $xM_2$, with $0.5<x<1$, at redshift $z$, is the integral 
quantity
\be
p(M>xM_2,z | M_2,z_2) = 
\int_{\ln (xM_2)}^{\ln M_2} d \ln M \left({M_2 \over M}\right)
{df \over d\ln M}
\label{eq:ftime}
\ee
with $z_2 = 0$, and $df/ d\ln M$ given by Eq.~(\ref{eq:cdmf}).
For $x=0.5$ LC94 found a good agreement between this prediction 
and their $N$-body simulations, for different cosmological models with
scale-free power spectrum. Here we compare our data with the same 
prediction, allowing however different values of $x$ between $0.5$ and
$1$, to consider a larger part of the cluster formation history.

\begin{figure}
\epsfxsize=\hsize\epsffile{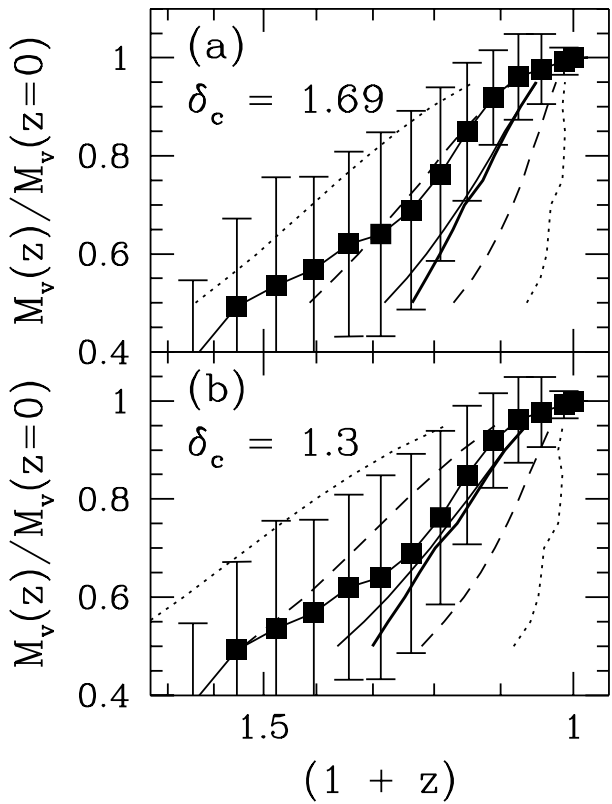}
\caption{Cluster formation redshift. Solid squares show the time
growth of the largest cluster progenitor, averaged over the
sample, and normalized to the final cluster mass. Error bars 
indicate $\pm 1\sigma$ of the distribution.
The other lines are obtained from Eq.~(\ref{eq:ftime}), and 
indicate the 5, 25, 50, 75 and 95 percentiles (dotted, dashed and thin
solid) and the mean (thick solid) of the predicted formation redhift 
for a cluster with mass $M_v = 1.1 \times 10^{15} M_{\sun}$, the 
sample mean mass. Panel {\em (a)} compares the data to the predictions 
for the calibrated \ps model ($\delta_c = 1.69$). 
Panel {\em (b)} is for $\delta_c = 1.3$.}
\label{fig:zform}
\end{figure}

Fig.~\ref{fig:zform} shows the probability distribution function 
in redshift for the mass of the main cluster progenitor. Symbols 
indicate the actual formation history of the largest progenitor,
normalized to the final cluster mass, and averaged over the cluster 
sample. Error bars give the $1\sigma$ dispersion of the sample 
distribution.
For each value of $x = M(z)/M_2(z=0)$, redshifts are indicated
which correspond to the mean (thick solid line), and to the 
5, 25, 50, 75 and 95 percentiles (dotted, dashed and thin solid 
lines) of the distribution obtained using Eq.~(\ref{eq:ftime}). 
The upper panel compares the data to the calibrated \ps model;
and the lower panel to the \ps model with $\delta_c = 1.3$.
The predictions are for a cluster with mass equal to the mean 
mass of the sample, $M_v = 1.1 \times 10^{15} M_{\sun}$.
Symbols above the mean indicate early formation, and viceversa.

The figure confirms the picture above: the sample clusters form
earlier than predicted by the calibrated \ps model.
For seven out of nine objects, the mass of the largest progenitor 
is always above the mean predicted at each redshift, while three 
clusters actually have a history at, or outside, the 95\% percentile 
of the predicted distribution.
For comparison, Fig.~\ref{fig:zform}b shows a \ps prediction with
$\delta_c = 1.3$, which provides a better description of the 
numerical data.

\subsection{Accretion rates}
\label{sec:arate}

We now consider the evolution of the halo infall onto the most massive
cluster progenitor. For clarity of terms, hereafter we will call the 
corresponding rate the cluster {\em accretion rate}, not to confuse 
it with the infall rate discussed in the previous Section.

Unlike the global infall rate, which can be calculated analytically,
the \ps prediction for the accretion rate onto the main progenitor 
requires the use of Monte Carlo merging trees, as the extended
\ps formalism does not contain, by itself, a rule on how to break
up the final cluster in smaller progenitors as one goes back in time.
To establish such a rule requires some extra assumptions, and
different assumptions lead to different ways to construct merging 
trees. However, although these may lead to differences in the details 
of the merging history, they should give similar global results 
(Kauffman \& White 1993).

The expression for the infall rate 
Eq.~(\ref{eq:irate}) does not distinguish between infall onto the 
most massive cluster progenitor and mergers between other progenitors. 
On the other hand, it is often useful to have the rate of infall
on the main progenitor of the cluster, especially if one wishes
to compare the latter to some observation. A classical example is 
the Butcher-Oemler effect (Butcher \& Oemler 1978): the fraction
of blue, star-forming galaxies in clusters increases rapidly with
redshift, so that in clusters at $z \sim 0.4$ it is a factor of 
$\sim 10$ higher than in clusters observed locally.

Two mechanisms have been advanced to give a physical explanation
to the starburst of these galaxies. The first is the infall hypothesis 
(Dressler \& Gunn 1983), where star formation is triggered by 
ram-pressure from the intracluster medium, during the first infall 
of an accreted galaxy in the cluster potential. 
A second idea (Lavery \& Henry 1988) is that galaxy-galaxy 
collisions in the cluster environment are responsible for the 
starburst.

Clusters formed in a hierarchical scenario have a history which 
is qualitatively consistent with both mechanisms (Bower 1991; 
Kauffmann 1995). In fact, a cluster of given physical mass is a 
much younger object at high than at low redshift. Therefore, both 
the infall of haloes and the merging of its progenitors are stronger 
at high redshift.
In these models, the quantity used to measure the infall of galaxies 
onto the cluster has usually been the global infall rate presented 
in Section~\ref{sec:irate} (Bower 1991), or the fraction of cluster 
mass in galaxy-size haloes (Kauffmann 1995).
However, the accretion rate onto the main cluster progenitor is 
probably a more appropriate quantity to consider. This differs
from the global infall rate especially early in the cluster
formation history (high lookback times), when most of the infall 
is between progenitors other than the most massive one. 

To highlight the difference between the accretion history of clusters 
observed at $z=0$ and at higher redshift, we divided our sample of
nine objects in two subsets: a first one containing the five less 
massive clusters, which we will observe at $z=0$ and will call 
{\em low redshift} (L) sample; and a second one with the four more 
massive clusters, which we will observe at $z\sim 0.55$ and will name 
{\em high redshift} (H) sample. As the average cluster mass in these 
two subsets at the time of observation is roughly the same, 
$M_v \approx 6 \times 10^{14} M_{\sun}$, we are looking at clusters 
of the same physical size at both redshifts, although the (H) clusters
are more massive in terms of $M_*$.

\begin{figure}
\epsfxsize=\hsize\epsffile{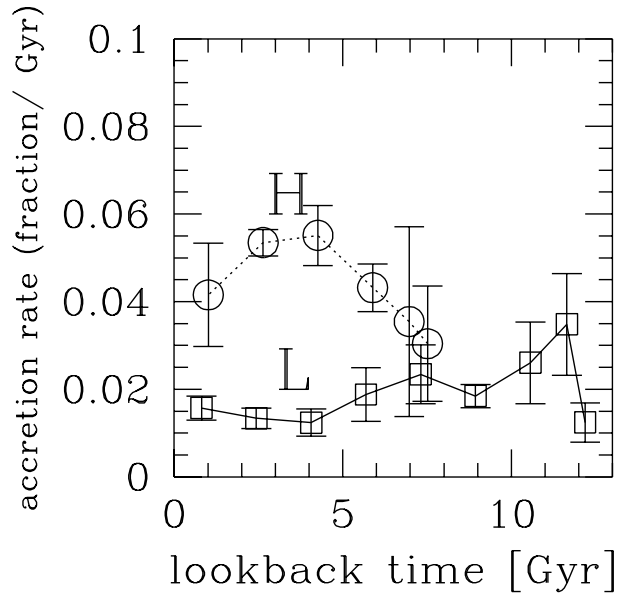}
\caption{Accretion rate onto the main cluster progenitor as a
function of lookback time. The figure shows the mass fraction 
accreted in haloes of mass $m_v\in[4 \times 10^{11},10^{13}] M_{\sun}$ 
per Gyr by clusters of average final mass $6 \times 10^{14}
M_{\sun}$. Squares correspond to the low redshift sample (L), 
circles to the high redshift sample (H). Error bars are $1\sigma$ 
of the sample mean.}
\label{fig:arate}
\end{figure}

\begin{figure}
\epsfxsize=\hsize\epsffile{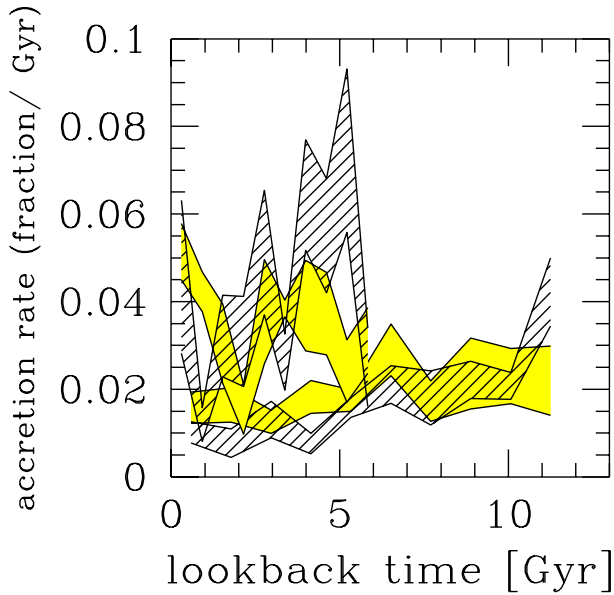}
\caption{The figure is the analog of Figure \ref{fig:arate}, but now 
plotting the prediction of the merger tree based on the extended
\ps formalism. The bands enclose $\pm 1\sigma$ (of the mean)
around the mean rate.}
\label{fig:aratet}
\end{figure}

Fig.~\ref{fig:arate} shows, for the two subsets, the evolution 
of the average accretion rate onto the main cluster progenitor,
as a function of lookback time. Only accretion of galaxy-size haloes 
is considered: $m_v \in [4 \times 10^{11}, 10^{13}] M_{\sun}$.
Squares show the accretion rate for low redshift (L) clusters, 
circles for (H) clusters. The corresponding global infall rate, 
for (L) clusters, is in-between that of the two upper panels of 
Fig.~\ref{fig:irate}. 
For the reason explained above, the (L) accretion rate is a much 
flatter function of lookback time than the global infall rate,
and is only slightly peaked at very high lookback times.
On he contrary, accretion in (H) clusters mostly happens very 
recently, when it is up to four or five times bigger than the
accretion rate for the (L) clusters.

Fig.~\ref{fig:aratet} shows the merging tree predictions to be
compared to Fig.~\ref{fig:arate}. The plotted bands enclose 
$\pm 1\sigma$ of the mean of the accretion rate on two samples of
clusters with the same characteristics as the two $N$-body subsets.
Solid bands are the prediction of the calibrated \ps model;
hatched bands are a \ps model with $\delta_c = 1.3$.  
As can be seen, accretion is stronger for the (H) sample, and 
weaker for the (L) sample, as in the $N$-body case. 
However, the predictions of different $\delta_c$ do not differ
appreciably. This is partly consequence of the large scatter in 
the predicted rates. However, the main reason for the similarity
is that here we consider accretion of galaxy-size haloes, and for
such haloes the two models predict similar abundances (as shown
in Fig.~\ref{fig:hab}) and have similar global infall rates
(from Fig.~\ref{fig:irate}). We therefore conclude that the accretion 
rate on the main cluster is not sensitive to the value of $\delta_c$.

Finally, Fig~\ref{fig:aratet} shows large variations in the accretion 
rates of the (H) sample as a function of lookback time, even within
the same model. This result suggests that even high redshift clusters
can sometimes exhibit, at the time of observation, a weak accretion 
rate, and that this is consistent with the discrete nature of matter 
infall. Such ``weakly accreting'' clusters could be naturally
associated to observations of ``non Butcher-Oemler'' clusters.

\section{Discussion and Conclusions}
\label{sec:concl}

We have studied the merging history of rich clusters, using $N$-body 
simulations of an Einstein-de Sitter universe, with scale-free power 
spectrum of fluctuations, and spectral index $n=-1$.
Our results are summarized as follows.
\begin{enumerate}
\item
The formation process of dark matter haloes in our simulations
satisfies the assumptions of the hierarchical clustering model 
with reasonable accuracy. 

\item
The global \ps mass function Eq.~(\ref{fig:udmf}) 
provides a good fit to $N$-body data. Although $df_M(M)/d\ln M$ 
from simulations is slightly {\em flatter} than predicted,
the standard choice $\delta_c=1.69$ produces a mass function which
agrees with the $N$-body data to better than 40 percent in the 
range $-1.4 \la \log_{10}(M/M_*) \la 1.4$, for both FOF(0.2) and
SO(178) haloes. This conclusion agrees with earlier work (e.g. 
Efstathiou et al. 1988; LC94).
A finer tuning in the value of the threshold density:  
$\delta_c=1.55$ for FOF(0.2) haloes and $\delta_c = 1.63$ for SO(178) 
haloes slightly extends the agreement at the high mass end of the
mass function.

\item
In contrast with the previous point, the assembly of matter in
a proto-cluster environment takes place earlier than predicted by 
the \ps model calibrated with $\delta_c=1.69$. As a consequence, 
this choice underestimates the growth of structures especially at 
the high mass tail and at high redshift. For example, the abundance 
of galaxy groups at redshift $\approx 5$ is underestimated by almost
two orders of magnitude. 

The discrepancy between theory and numerical data can be modeled
by a simple variation of the collapse threshold: a value 
$\delta_c=1.3$, which corresponds to a typical nonlinear mass 
$M_*$ more than two times larger than that of the calibrated
\ps model, would give a much better fit to the $N$-body data.
However, $\delta_c$ was fixed at $1.69$ by the requirement that
the \ps model best-fits the global mass function, and is not a 
free parameter anymore.

\item
The accretion rate of galaxy-size haloes onto the main cluster
progenitor is not sensitive to the choice of $\delta_c$, as 
it samples a mass range where the difference between the two
considered models are not significative.
\end{enumerate}

The fact that the \ps model, calibrated to fit the global mas
function, is unable to describe the clustering of a constrained 
environment suggests that the \ps model and extensions, in their 
present formulation, may be structurally inadequate to give a 
consistent description of the gravitational clustering of matter 
in a general context. 
Such a conclusions would have important theoretical consequences 
if confirmed by future work. 

We stress that the present result has nothing to do with the 
different values of $\delta_c$ found by several authors when
fitting the global mass function (e.g. Efstathiou et al. 1988;
Gelb \& Bertschinger 1994; LC94). 
As clearly dicussed by LC94, there the discrepancies were mainly 
due to different choices for the algorithm used to define dark matter 
haloes and for the filter used to define the linear prediction.
On the contrary, we find that the value of $\delta_c$ that best-fits
the global mass function is inadequate to fit the statistics
of halo progenitors which come from the same simulations, and are
defined using the same group-finding algorithm and linear filter. 
Such comparison has never been done before and highlights a
novel weakness of the \ps model.

One possibile interpretation of the present results is that the 
value of the collapse threshold appropriate for the conditional 
statistics might depend on the mass of the final object relative 
to the value of $M_*$. In this interpretation, $\delta_c$ would 
be regarded as a free parameter to be fixed as a function of mass.
The value we found, $\delta_c = 1.3$, would then be suitable for 
rich clusters, i.e. objects with $M \ga 10 M_*$. Higher values 
might instead be required to fit the conditional statistics of smaller 
haloes, so that the standard threshold collapse value, $\delta_c=1.69$, 
would result from the {\em average} of a continuous set of different 
values over the entire halo population.

A physical explanation of such a dependence could be the departure
from spherical collapse (Bond \& Myers 1996; Monaco 1995) in a
mass-dependent manner; another could be some systematic difference
in the properties of the initial conditions for objects of
different mass. 
To test the first possibility, we measured the infall pattern of 
matter for the clusters in our sample, using the axial ratios of 
the ellipsoid defined by satellite infall onto the main cluster 
progenitor (as in Tormen 1997). We did not find any systematic 
difference between high and low mass clusters; however our sample 
is too small in number and in mass range to make a definitive 
statement on this point. 
The second possibility requires a more targeted analysis, again
using a larger cluster sample and spanning a larger mass range, 
and work is in progress in this direction.

In the light of the present results, we conclude by stressing that 
the \ps model should be used with care when describing the clustering 
properties of halo progenitors, as such description, in its standard
formulation, is not accurate.

\section*{ACKNOWLEDGEMENTS}

I would like to thank Simon White for his valuable comments 
and suggestions to the present work, and for kindly providing 
the cosmological simulations used for the analysis of Section 3.
Many thanks also to Shaun Cole, Antonaldo Diaferio, Bhuvnesh Jain, 
Cedric Lacey, Sabino Matarrese, Lauro Moscardini, Julio Navarro and 
Ravi Sheth for helpful discussions and comments.
Thanks to Ravi Sheth also for providing the merger tree data
used to produce Fig.~\ref{fig:aratet}.
Financial support was provided by an MPA guest fellowship and 
by the Training and Mobility of Researchers European Network 
``Galaxy Formation and Evolution''.
The simulations were performed at the Institut d'Astrophysique de
Paris, which is gratefully acknowledged.

\end{document}